\documentclass[prb,twocolumn,showpacs,preprintnumbers,amsmath,amssymb,superscriptaddress]{revtex4-1}

\usepackage{amsmath,amssymb,amsfonts,mathrsfs} 	
\usepackage{graphicx}
\usepackage{subfigure}

\renewcommand{\[}{\begin{equation}}
\renewcommand{\]}{\end{equation}}
\def\bea{\begin{eqnarray}}
\def\eea{\end{eqnarray}}
\def\nn{\nonumber\\}

\newcommand{\emi}[1]{{\rm e}^{-i #1}}
\newcommand{\ei}[1]{{\rm e}^{i #1}}

\newcommand{\intk}{\int_{\rm BZ} \!\! \frac{d {\bf k}}{(2\pi)^d} \;}
\newcommand{\E}{{\bf E}}

\renewcommand{\j}{{\bf j}}

\newcommand{\q}{{\bf k}}
\renewcommand{\k}{{\bf k}}

\renewcommand{\v}{{\bf v}}
\renewcommand{\r}{{\bf r}}

\newcommand{\da}{\partial_{k_\alpha}}
\newcommand{\db}{\partial_{k_\beta}}

\newcommand{\equ}[1]{Eq.~(\ref{#1})}

\def\ket#1{\vert#1\rangle}

\def\me#1#2#3{\langle#1| \, #2 \, |#3\rangle}
\def\runtime{(\the\time)\qquad\the\month/\the\day/\the\year}
\def\today
 {\count10=\year\advance\count10 by -2000 \number\day--\ifcase
  \month \or Jan\or Feb\or Mar\or Apr\or May\or Jun\or
             Jul\or Aug\or Sep\or Oct\or Nov\or Dec\fi--\number\count10}

\def\hour{\count10=\time\count11=\count10
\divide\count10 by 60 \count12=\count10
\multiply\count12 by 60 \advance\count11 by -\count12\count12=0
\number\count10 :\ifnum\count11 < 10 \number\count12\fi\number\count11}

\begin{document}

\title{Drude weight in systems with open boundary conditions}

\author{Gabriele Bellomia} 
\affiliation{International School for Advanced Studies (SISSA), Via Bonomea 265, 34136 Trieste, Italy} \email{gabriele.bellomia@sissa.it}
\author{Raffaele Resta}
\affiliation{Istituto Officina dei Materiali IOM-CNR, Strada Costiera 11, 34151 Trieste, Italy} \email{resta@iom.cnr.it}
\affiliation{Donostia International Physics Center, 20018 San Sebasti{\'a}n, Spain}

\begin{abstract}
A many-electron conducting system undergoes free acceleration in response to a macroscopic field. The Drude weight $D$---also called charge stiffness---measures the adiabatic (inverse) inertia of the electrons; the $D$ formal expression requires periodic boundary conditions. When instead a bounded sample is addressed within open boundary conditions, no current flows and a constant (external) field only polarizes the sample: the Faraday cage effect. Nonetheless a low-frequency field induces forced oscillations: we show here that the low-frequency linear response of the bounded system is dominated by the adiabatic inertia and allows an alternative evaluation of $D$. Simulations on model one-dimensional systems demonstrate our main message.
\end{abstract}


\maketitle \bigskip\bigskip

\section{Introduction}

Irrelevance of the boundary conditions in the thermodynamic limit is a basic tenet of statistical mechanics and condensed matter physics. Among the possible choices of boundary conditions two are prominent: Born-von-K\`arm\`an periodic boundary conditions (PBCs) and the so-called ``open'' boundary conditions (OBCs). Insofar as an intensive physical observable is computed from {\it finite} realizations of a given system, the two choices yield somewhat different results. Yet one postulates that the large-system limit yields the same value for any intensive physical observable.

To be more specific, we will consider below the ground state of a macroscopically homogeneous system of $N$ electrons and a neutralizing background of nuclei in a cubic box of volume $L^d$ ($d$ is the dimension). The choice of PBCs~vs.~OBCs amounts to choosing two different Hilbert spaces for describing our system: within PBCs the many-body wavefunction is periodic with period $L$ over each Cartesian coordinate of each electron independently, while within OBCs it is required to vanish whenever a coordinate is outside the box.

Some intensive physical observables are non problematic: this is e.g.~the case of spectral properties. At finite size the spectra are discrete within both  OBCs and PBCs, and different between themselves. In the large-system limit the two spectra become continuous and coincide, yielding the same density of states. Indeed, it is a standard exercise to verify this in the special case of a free-electron gas, which can be worked out analytically. Some other properties are more problematic, and were understood relatively recently: in this class are electrical polarization and orbital magnetization---they are trivial within OBCs and highly nontrivial within PBCs;  for a thorough analysis of both observables see e.g.~Ref. \onlinecite{Vanderbilt}.

The Drude weight $D$ (also called charge stiffness)\cite{Kohn64,Kohn68,Shastry90,Scalapino93,Allen06} measures the effective density-to-mass ratio contributing to dc electronic conductivity. A milestone paper by Kohn, formulated within PBCs, provided in 1964 the most general definition for $D$ in any macroscopically homogeneous system, including cases with disorder and electron-electron interaction.
Within OBCs, instead, $D$ apparently does not exist, given that a bounded sample does not support dc currents; the apparent paradox was previously addressed in Ref.~\onlinecite{Rigol08}, where it was indeed shown that an accurate treatment of the thermodynamic limit allows retrieving the $D$ value even from OBC simulations. 
While Ref.~\onlinecite{Rigol08} is rooted in lattice models (with and without interaction, at very high temperature), here we adopt the framework of zero-temperature electronic structure, by addressing noninteracting electrons in a periodic potential (in a mean-field sense). Our variational simulations highlight the relevance of the $f$-sum rule, whose accurate fulfillment requires near completeness of the basis set, as the key quantity to control the numerical error on our results. Most important, our choice of the case study naturally yields a clear physical interpretation: the effective density-to-mass ratio of the many-electron system can be accurately probed in two different ways: either via the response of an unbounded system to a constant field, or via the response of a bounded crystallite to a low-frequency oscillating field. In the former case one probes free acceleration, in the latter forced oscillations.

\vfill

In Sec.~II we provide the main definitions and we parse the $f$-sum rule; in Sec.~III we display the $D$ formal expressions within band-structure theory. The Kubo formula for conductivity at the independent-particle level, within both PBCs and OBCs, is presented in Sec.~IV; the differences between the two cases are thoroughly discussed. The Souza-Wilkens-Martin sum rule,\cite{Souza00} a very powerful tool to discriminate insulators from metals, is presented in Sec.~V. The results of our one-dimensional (1$d$) simulations are presented and discussed in Sec.~VI; we address separately free electrons, band insulators, and the most relevant case of band metals, which perspicuously demonstrates our major claim. In Sec.~VII we draw some conclusions, and in the Appendix we analyze in some detail the relationship between Kohn's approach and Kubo formul\ae.

\vfill

\section{Phenomenology}

The conductivity tensor $\sigma_{\alpha\beta}(\omega)$ yields the current density  linearly induced by a macroscopic electric field at frequency $\omega$ (Greek subscripts are Cartesian indices); for the sake of simplicity we assume time-reversal symmetry, in which case the transverse conductivity vanishes and $\sigma_{\alpha\beta}(\omega)$ is a symmetric tensor.

In a metal, in the absence of dissipation, the electrons in a dc field undergo free acceleration and 
$\sigma_{\alpha\beta}(\omega)$ is divergent for $\omega=0$. The most general form for longitudinal conductivity is\cite{Shastry90,Scalapino93,Allen06} \[ \sigma_{\alpha\beta}(\omega) = D_{\alpha\beta} \left[ \delta(\omega) + \frac{i}{\pi \omega} \right] +\sigma_{\alpha\beta}^{(\rm regular)}(\omega) , \label{cond} \] where the constant $D_{\alpha\beta}$ goes under the name of Drude weight (or charge stiffness) and accounts for the inertia of the many-electron system in the adiabatic limit.\cite{Scalapino93,rap157} The Drude weight can also be defined as\cite{Kohn64} 
\[ D_{\alpha\beta} = \pi \lim_{\omega \rightarrow 0} \omega \; \mbox{Im } \sigma_{\alpha\beta}(\omega) \label{imlim}  . \]
Longitudinal conductivity obeys the the $f$-sum rule, \bea \int_0^\infty d \omega \; \mbox{Re } \sigma_{\alpha\beta} (\omega) &=& \frac{D_{\alpha\beta}}{2} + \int_0^\infty d \omega \; \mbox{Re } \sigma_{\alpha\beta}^{(\rm regular)} (\omega) \nn &=& \frac{\omega_{\rm p}^2}{8}\delta_{\alpha\beta} = \frac{\pi e^2 n}{2 m}\delta_{\alpha\beta} , \label{fsum} \eea where $n=N/L^d$ is the electron density and $\omega_{\rm p}$ is the plasma frequency. For free electrons $\sigma_{\alpha\beta}^{(\rm regular)}(\omega)$ vanishes and  $D_{\alpha\beta}$ assumes the same value as in classical physics,\cite{Drude00,AM1} i.e $D_{\alpha\beta} = D_{\rm free} \,\delta_{\alpha\beta}$, with $D_{\rm free} = \pi e^2 n/m$.  Given \equ{fsum}, switching the periodic potential on has the effect of transferring some spectral weight from the Drude peak into the regular term; for band insulators the Drude peak vanishes and $\mbox{Re }\sigma_{\alpha\beta}^{(\rm regular)} (\omega)$ is zero for $\omega < \epsilon_{\rm gap}/\hbar$.
In the special case of a band metal considered here $\sigma_{\alpha\beta}^{(\rm regular)}(\omega)$ is a linear-response property which accounts for interband transitions, and is nonvanishing only at frequencies higher than a finite threshold; in the more general case of a noncrystalline many-electron system this selection rule breaks down and $\sigma_{\alpha\beta}^{(\rm regular)}(0)$ may be nonzero.\cite{Scalapino93} 

\section{Drude weight}

When applied to a band metal with doubly occupied orbitals, within PBCs, Kohn's general expression\cite{Kohn64,Kohn68,Scalapino93} becomes the Fermi-volume integral\cite{Allen06} \[ D_{\alpha\beta} = 2 \pi e^2 \sum_{j} \intk \theta(\mu - \epsilon_{j\k}) \, m^{-1}_{j,\alpha\beta}(\k) , \label{band} \]  where BZ is the Brillouin zone, $\mu$ is the Fermi level, and the effective inverse  mass tensor of band $j$ is  \[ m^{-1}_{j,\alpha\beta}(\k) = \frac{1}{\hbar^2} \frac{\partial^2 \epsilon_{j\k}}{\partial k_\alpha \partial k_\beta} . \label{mass} \] 
For insulators, the integral in \equ{band} trivially vanishes; for metals, the contribution of the core bands to $D_{\alpha\beta}$ vanishes as well. $D_{\alpha\beta}$ can be equivalently expressed as a Fermi-surface integral, by means of an integration by parts; it acquires then the meaning of an ``intraband'' term:\cite{Allen06} \[  D_{\alpha\beta} = - 2 \pi e^2 \sum_j \intk  f'(\epsilon_{j\k}) \; v_{j \alpha}(\k) v_{j \beta}(\k) , \label{Allen1} \] where $v_{j\alpha}(\k) =  \da \epsilon_{j\k} / \hbar $ and  at zero temperature the Fermi occupation function is $f(\epsilon) = \theta(\mu - \epsilon)$. \equ{Allen1} is in explicit agreement with the spirit of Landau’s Fermi-liquid theory, which holds that charge transport in metals involves only quasiparticles with energies within $k_{\rm B}T$ of the Fermi level; \equ{Allen1} is in fact at the root of the semiclassical theory of transport.\cite{AM-Ch13}

Notice that so far we have not explicitly invoked the Kubo formul\ae\ for conductivity; this is a virtue of Kohn's approach, where they remain implicit. The above results can be equivalently formulated via Kubo formul\ae\ at ${\bf q}=0$ (not ${\bf q} \rightarrow 0$);\cite{nota} furthermore the scalar potential of a constant field is incompatible with PBCs,\cite{rap100} where it is mandatory to adopt the vector-potential gauge instead.\cite{Shastry90,Scalapino93,rap157} Some more details about the relationship between Kohn's formula, \equ{band}, and the equivalent sum-over-states Kubo formula for $D$ are given in the Appendix; the many-body analog can be found in Ref.~\onlinecite{rap157}.

\section{Kubo formula} \label{sec:kubo}

The Kubo formul\ae\ can be cast in several equivalent ways; here it is expedient to adopt the form\cite{Allen06} \[ \sigma_{\alpha\beta} (\omega) = \frac{2 i e^2 \hbar}{L^d} \sum_{mn} \left(\frac{f_n - f_m}{\epsilon_m - \epsilon_n} \right) \frac{\me{n}{v_\alpha}{m} \me{m}{v_\beta}{n}}{\hbar(\omega + i\eta)+\epsilon_n - \epsilon_m} , \label{general} \] where  the velocity is $\v = i[H,\r]/\hbar$, the positive infinitesimal $\eta$ enforces causality, and $f_n = 1/({\rm e}^{\beta \epsilon_n} + 1)$ is the Fermi occupation factor.

The previous Eqs.~(\ref{band})-(\ref{general}) by definition address solely noninteracting electrons, where pairing and superconductivity are ruled out. It is worth noticing that the corresponding Kubo formul\ae\ for interacting electrons may comprise an extra term, accounting for superconducting (a.k.a.~Meissner) weight.\cite{Shastry90,Scalapino93,rap157}

The independent-electron expression of \equ{general} holds both within OBCs and PBCs. 
When a bounded crystallite (cut from a bulk metal) is addressed within OBCs, \equ{general} does not account for a Drude peak at finite size; it also follows that the $\omega>0$ region of the spectrum saturates the $f$-sum rule.\cite{Akkermans97}

When instead PBCs are adopted the index $n$ must be identified with the band and Bloch index $j\k$. Owing to the Bloch theorem, it is possible to perform 
the thermodynamic limit first---from discrete to continuous $\k$---where the diagonal elements $\me{n}{\v}{n}$ are identified with $\v_j(\k) = \partial_{\k} \epsilon_{j\k}/\hbar$, and the factor $(f_n-f_m)/(\epsilon_m-\epsilon_n)$ with $-f'(\epsilon_{j\k})$. 
In this case the $T \rightarrow 0$ limit of \equ{general} yields---besides the interband (regular) term---the additional intraband (Drude) term \[  \sigma_{\alpha\beta}^{(\rm intra)}(\omega) = - \frac{2e^2}{L^d}\frac{i}{\omega+i \eta}\sum_{j\k} f'(\epsilon_{j\k}) \; v_{j \alpha}(\k) v_{j \beta}(\k) , \label{agree} \] where the $\k$-sum is actually an integral, \[ \frac{1}{L^d}  \sum_{j\k} \rightarrow \sum_{j} \intk . \label{limit} \]
Therefore \equ{agree} clearly coincides with \equ{Allen1}; adding this term to the regular one the $f$-sum rule is retrieved.

Matters are different if one performs the $T \rightarrow 0$ limit first, as we are going to do here in order to compare PBC and OBC expressions on the same ground at finite size: in both cases all levels are discrete. \equ{general} yields, 
for an isotropic system and at $\omega>0$, \[ \mbox{Re } \sigma(\omega) = \frac{2 \pi e^2}{\hbar L^d} \sum_{\substack{\epsilon_n \leq \mu \\ \epsilon_m > \mu}} \frac{|\me{n}{v_x}{m}|^2}{\omega_{nm}} \delta(\omega - \omega_{nm}) , \label{Allen2}
\] where $\omega_{nm} = (\epsilon_m - \epsilon_n)/\hbar$. \equ{Allen2} obeys the $f$-sum rule in the OBCs case, but instead {\it does not} saturate it in the PBCs metallic case:\cite{Akkermans97} in fact \equ{Allen2} in the thermodynamic limit yields the regular term  $\sigma_{\alpha\beta}^{(\rm regular)}(\omega)$ only, and the $f$-sum rule is satisfied only if the $D$ value, evaluated from Kohn's formula, is added.\cite{Shastry90}
We stress that the matrix elements and the selection rules are quite different in the PBCs vs.~OBCs cases. In the special case of free electrons the PBCs orbitals are plane waves and all matrix elements in \equ{Allen2} vanish because of an obvious selection rule. 

At any finite size all poles in \equ{Allen2} occur at positive energies, within both PBCs and OBCs. There is an outstanding difference, though: the PBCs poles are gapped, while a subset of the OBCs poles converge to zero frequency. The selection rules forbid intraband PBCs contributions to \equ{Allen2}, while instead within OBCs the intraband transitions originate low-frequency poles, which contribute with extra spectral weight to the $f$-sum rule. In agreement with the previous findings of Ref.~\onlinecite{Rigol08}, the most relevant of the low-energy poles coalesce, for $L \rightarrow \infty$, into a single pole at $\omega=0$, whose residue yields $D$. In the following we are going to study this process in detail for a few one-dimensional test cases. 

\section{Souza-Wilkens-Martin sum rule}

Souza, Wilkens, and Martin (hereafter referred to as SWM)\cite{Souza00} proposed in 2000 to characterize the metallic/insulating behavior of a material by means of the integral (for isotropic systems) \[ I^{(\rm SWM)} = \int_0^\infty \frac{d \omega}{\omega} \mbox{Re } \sigma(\omega) \label{swm} , \] which diverges for all metals and converges for all insulators. We adopt here the SWM approach, but we stress that---at finite size---its PBC features are quite different from the OBC ones.

In a band metal $I^{(\rm SWM)}$ diverges within PBCs because of the $\delta$-like Drude peak, which exists even at finite size; equivalently, it diverges because a dc field induces free acceleration (again, even at finite size).
Within OBCs, instead, all of the poles of $\sigma(\omega)$ occur at nonzero frequency; $I^{(\rm SWM)}$ is finite at any size, and diverges in the large-system limit. Our simulations will show that such a divergence is due to the low-frequency poles which are the fingerprint of $D$ within OBCs: the system cannot undergo free acceleration, but when the size is increased the forced oscillations decrease in energy and couple to the field with nonvanishing oscillator strength.

In a band insulator $I^{(\rm SWM)}$ is finite both within PBCs and OBCs, while we expect the integrated values to converge towards the same large-system limit. From \equ{Allen2} we get the sum \[ I^{(\rm SWM)} = \frac{2 \pi e^2}{\hbar L^d} \sum_{\substack{\epsilon_n \leq \mu \\ \epsilon_m > \mu}} \frac{|\me{n}{v_x}{m}|^2}{\omega^2_{nm}} ; \label{swm2} \] when evaluated within PBCs vs.~OBCs its terms differ in energies, matrix elements, and selection rules. The lowest PBC transition energy is gapped, while no selection rule forbids low-energy transitions within OBCs: this fact is at the root of the divergence of the OBC SWM integral, \equ{swm2}, in the metallic case.

By exploiting completeness, $I^{(\rm SWM)}$ can be transformed into a ground-state property: therefore the organization of the electrons in the ground state---and {\it not} a spectral gap---discriminates in general between insulators and metals.\cite{Kohn64,Souza00,rap132,rap_a33,rap156} Such behavior has been understood so far in terms of the mean-square quantum fluctuation of the polarization within OBCs; here we provide a different interpretation in terms of the Drude weight.

\section{Simulations}

From now on we address $D$ in units of $\pi e^2 n/(2m)= \omega_{\rm p}^2/8$, such that the $f$-sum rule yields $1$. Therefore 
$D_{\rm free}/2=1$ for free electrons, and $D/2 < 1$ for a generic metal; the conductivity will be displayed in units of $\omega_{\rm p}/2$ throughout.

\subsection{Free electrons}

\begin{figure}[t] 
~~\\ \bigskip 
\includegraphics[width=0.98\linewidth]{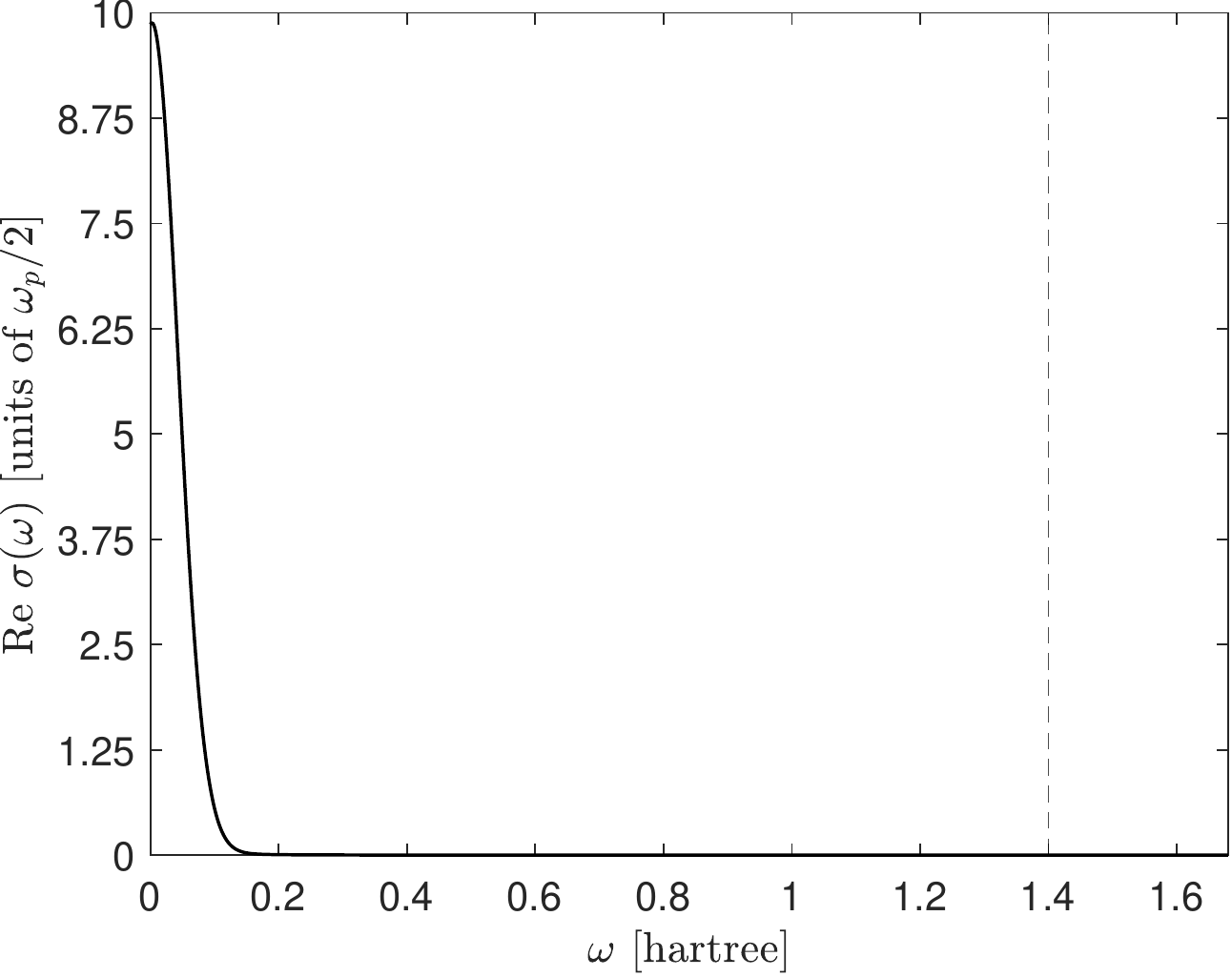} \\ 
\caption{Free-electron conductivity in units of $\omega_{\rm p}/2$ for the infinite one-dimensional potential well for $n=N/L=0.2$ bohr$^{-1}$ and $N=162$. The cutoff is also shown (vertical dashed line).
\label{fig:sigma-free}} 
\end{figure}

As said above, the free-electron case in PBCs is trivial: $\sigma^{(\rm regular)}(\omega) \equiv 0$ and all of the spectral weight goes into the $\delta(\omega)$ term. In the OBCs case we have computed \equ{Allen2} using the (analytical) eigenvalues and eigenfunctions of a 1$d$ infinite potential well of length $L$, at a linear density $n=N/L=0.2$ bohr$^{-1}$, with double orbital occupancy; the series has been truncated by means of a cutoff in the included excitation energies. The result is shown in Fig.~\ref{fig:sigma-free} for $N=162$ and a cutoff of 1.4~Ha; the $\delta$-singularities have been plotted, as customarily, as narrow Gaussians. The figure perspicuously show that the poles of \equ{Allen2} accumulate at very low energy; the value of $\sigma(0)$ does not carry any physical meaning, since it depends on the (arbitrary) Gaussian smearing. We emphasize instead the accurate integrated (smearing-independent) value of $\sigma(\omega)$: the $f$-sum rule is satisfied here at 99.99\% of the exact value.

\vfill

\begin{figure}[t] 
\includegraphics[width=0.84\linewidth]{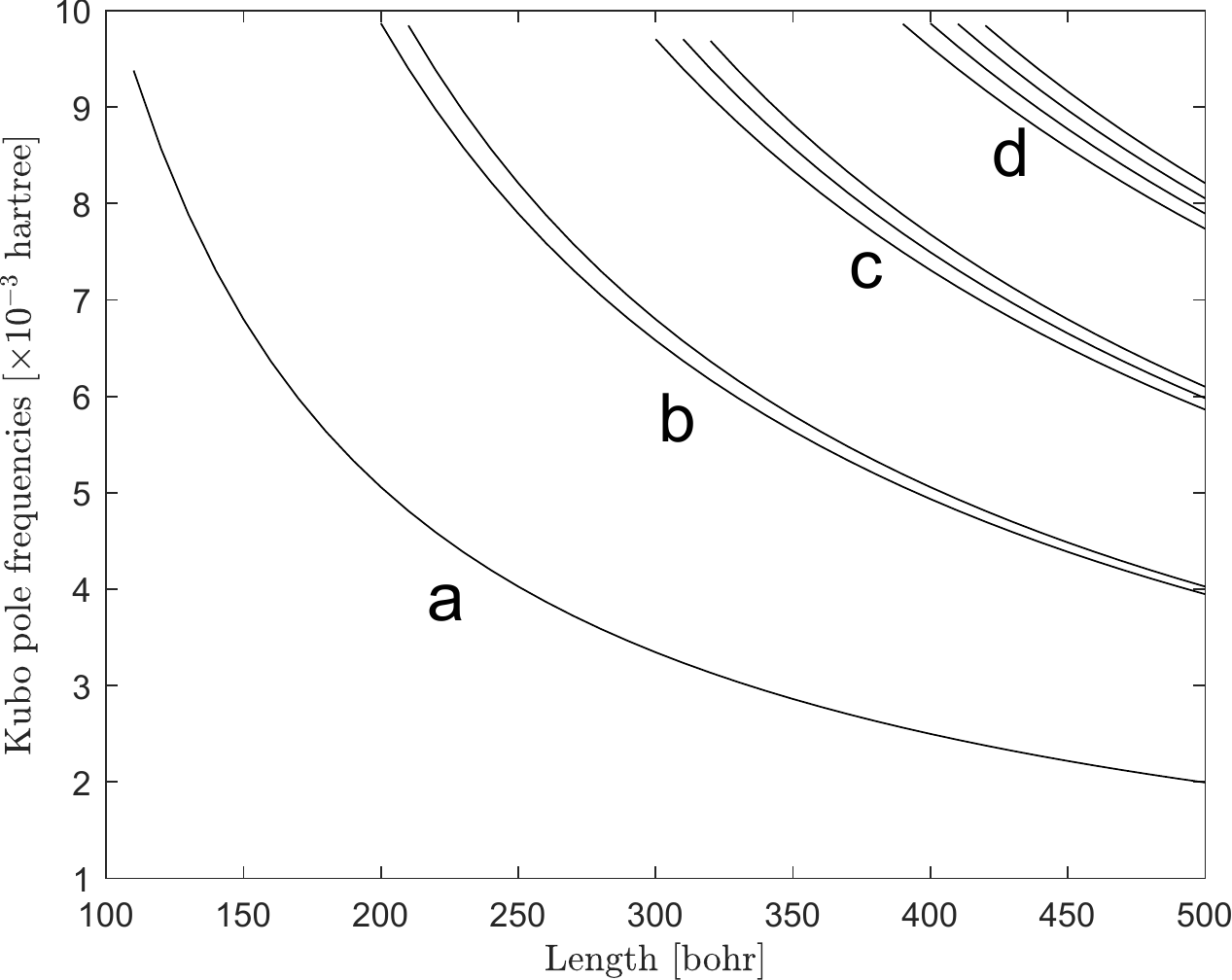} \\ \bigskip
\includegraphics[width=0.85\linewidth]{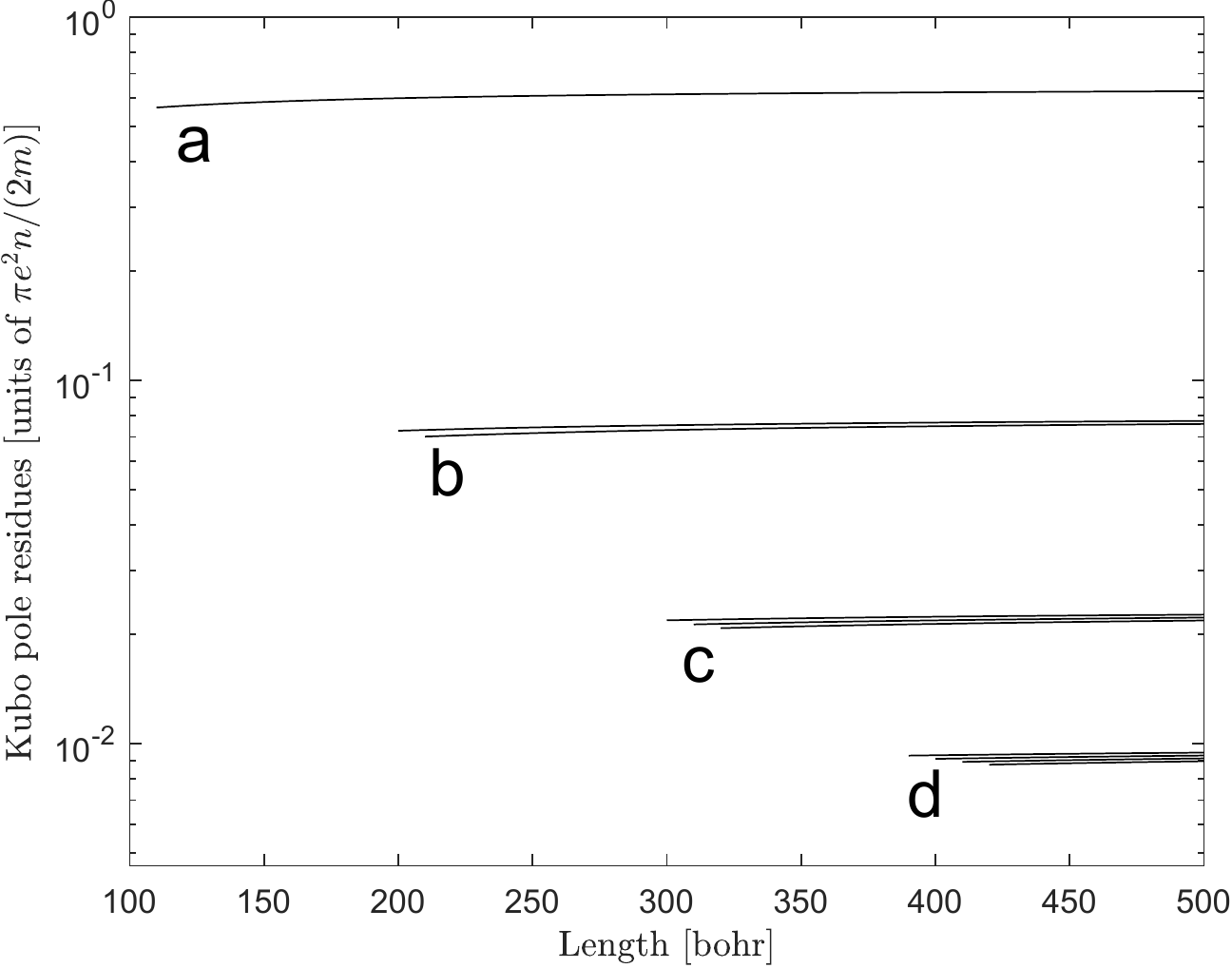}
\caption{$L$-dependence of the poles in \equ{Allen2}, at a low cutoff $\epsilon_{\rm cut} = 0.01$ Ha. The top panel shows the pole frequencies: the number of poles increases with $L$, there are families of poles, each family following a $1/L$ law. The bottom panel shows the corresponding residues in units of $\pi e^2 n/(2m)= \omega_{\rm p}^2/8$, exponentially vanishing with the family index.
\label{fig:poles}} 
\end{figure}

From Fig.~\ref{fig:sigma-free} it is seen that the poles carrying non-negligible residues occur at frequencies lower than about 0.15 Ha; this value is clearly size-dependent, since the spacing of the levels goes as $1/L$. One would expect that all poles converge towards zero in the large-$L$ limit; but the situation, illustrated in Fig.~\ref{fig:poles}, is much less trivial. At any given cutoff, there are several families of poles, whose number increases with size. Within a given family the frequency follows a $1/L$ law, as shown in Fig.~\ref{fig:poles} (top panel); the cutoff has been lowered with respect to Fig.~\ref{fig:sigma-free} for the sake of clarity. We also find (Fig.~\ref{fig:poles}, bottom) that the pole residues are essentially $L$-independent and that they are exponentially vanishing with the family index; therefore only a small number of low-frequency poles carry significant residues. The message of both panels altogether is therefore that, despite a complex pole pattern, the spectral weight is confined in a frequency region proportional to $1/L$.
  
\vfill
  
\begin{figure}[t]
\centering
\begin{minipage}[b]{.50\linewidth}
\includegraphics[width=\linewidth]{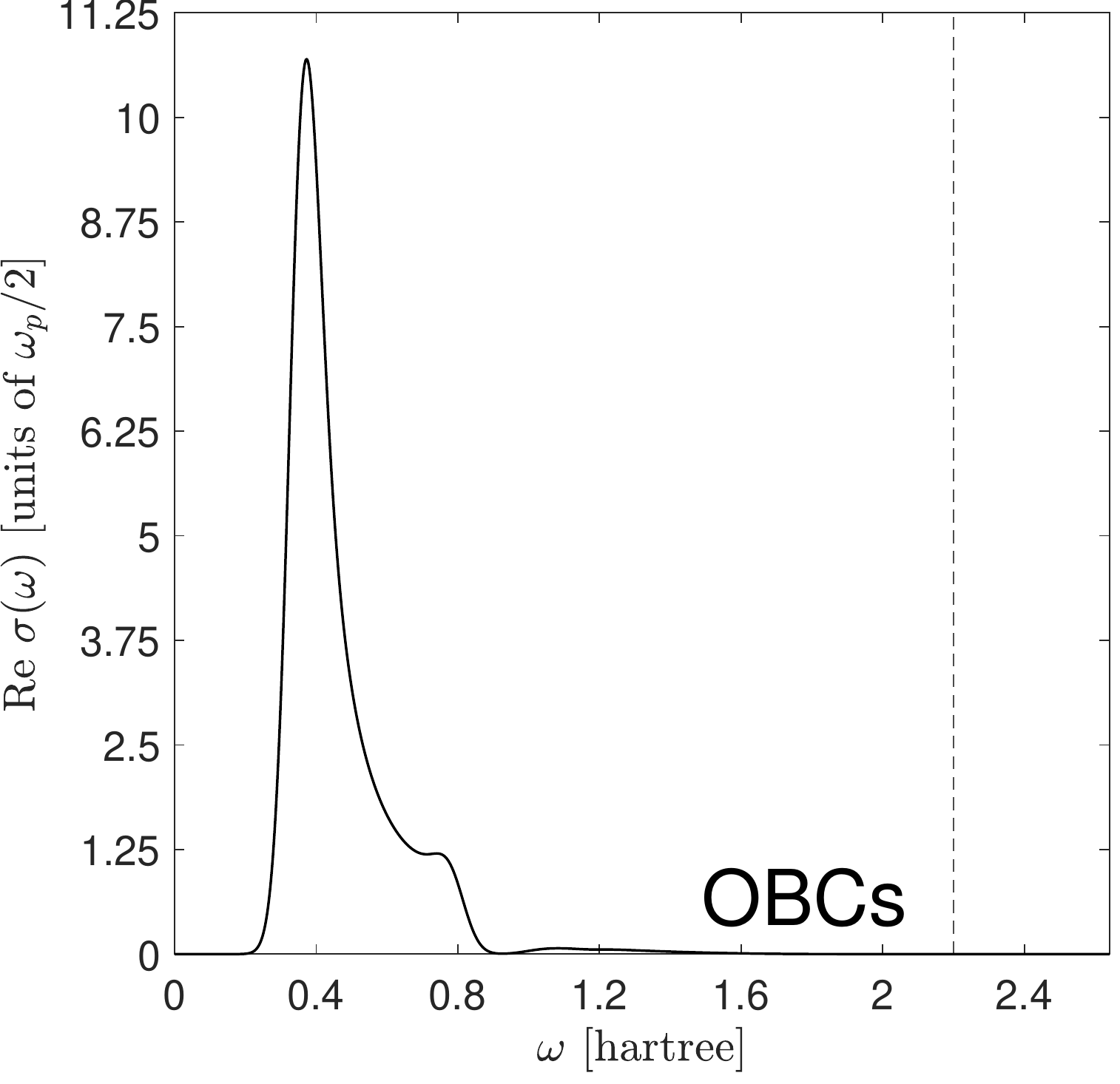}
\end{minipage}\hfill
\begin{minipage}[b]{.50\linewidth}
\includegraphics[width=\linewidth]{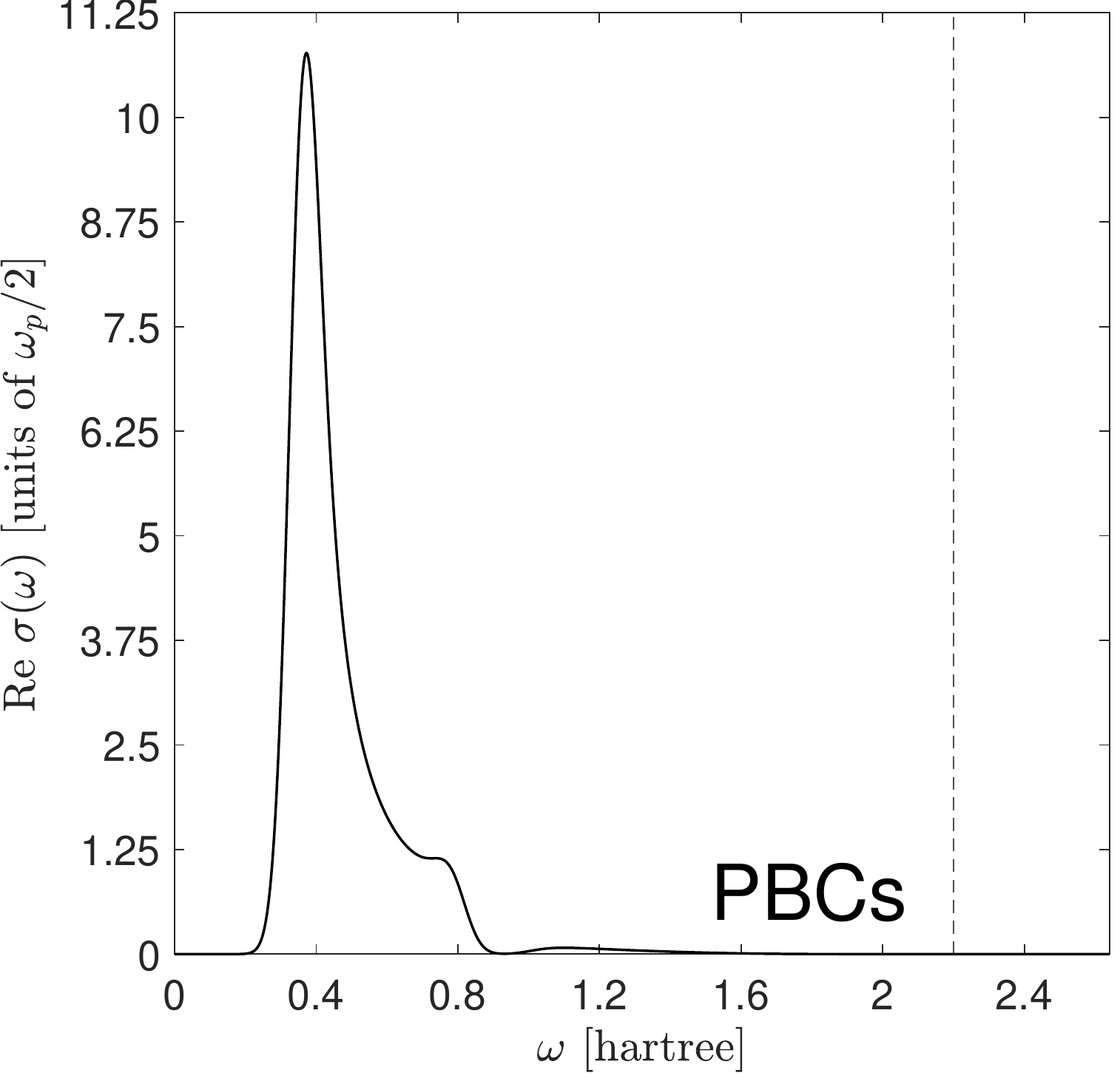}
\end{minipage}
\caption{Conductivity of the model insulator in units of $\omega_{\rm p}/2$,  after \equ{Allen2}. The gap is $\epsilon_{\rm gap} = 0.35$ Ha; the energy cutoff is shown as a vertical dashed line. Left panel: OBCs. Right panel: PBCs.
}
\label{fig:ins} \end{figure}

\subsection{Periodic potential}

Next we switch on a potential in the form of a periodic array of Gaussians: \[ U(x) = \sum_{m\,=-\infty}^\infty V(x-ma), \quad V(x) = V_0 \,\,{\rm e}^{-x^2/b^2} ; \] we set $a=5$  and $b=1$ bohr. We get a model metal with 1 electron/cell and a model insulator with 2 electrons/cell; in the former case the density is the same as for the free-electron case, discussed above. The eigenproblem is solved here numerically by representing the solutions on a basis of 700 free electron states: plane-waves for the PBCs ring and stationary sine-waves for the OBCs infinite potential well. By choosing $V_0 = 0.8$ Ha the first gap in the spectrum is $\epsilon_{\rm gap} = 0.35$ Ha. The excitations in the Kubo formul\ae~are again included up to a given energy cutoff: 2.2~Ha for both fillings.

\subsection{Model insulator}

\begin{figure}[t] 
\includegraphics[width=0.80\linewidth]{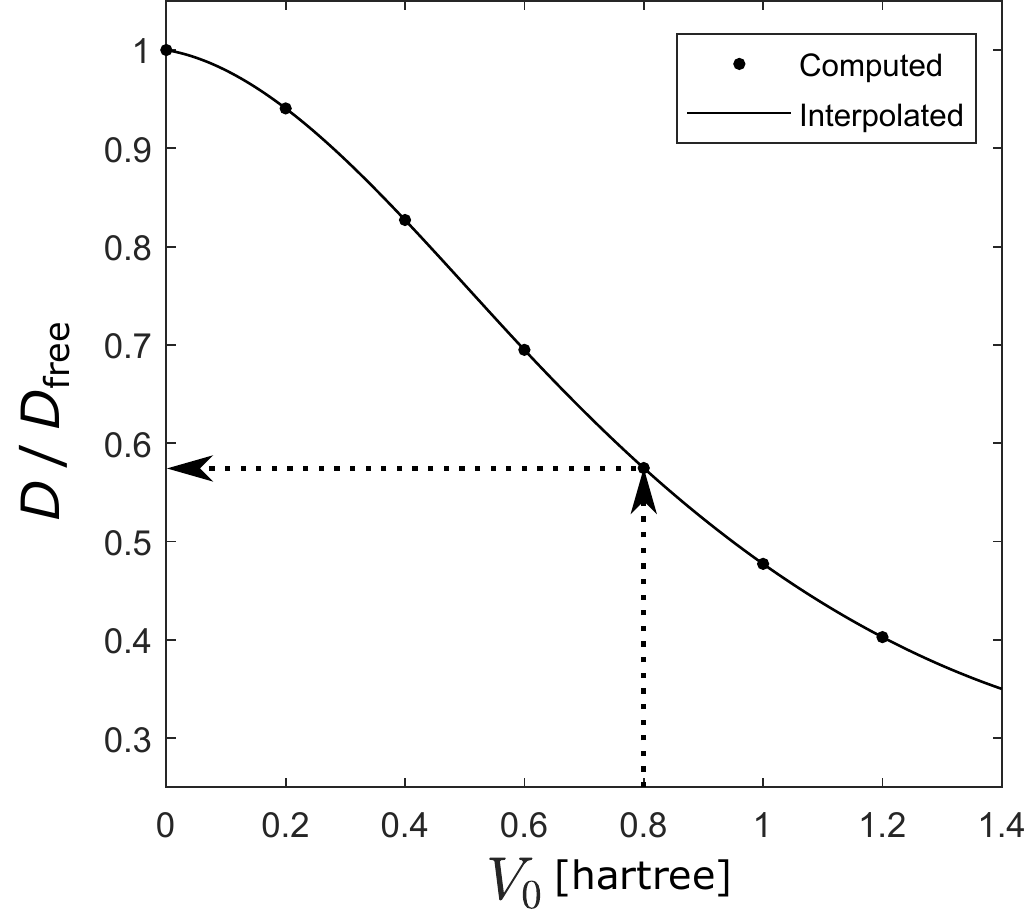} \\ 
\caption{Drude weight as a function of the periodic potential strength $V_0$.
\label{fig:strength}} 
\end{figure} 

We start with showing the results of the (almost trivial) insulating case. With $V_0 = 0.8$ Ha and a cutoff of 2.2 Ha we are very close to completeness: $f$-sum fulfilled at 99.93\% and 99.99\% within OBCs and PBCs, respectively. The conductivity plots evaluated from \equ{Allen2} in the two cases are basically undistinguishable (Fig.~\ref{fig:ins}); the SWM integrals, \equ{swm2}, differ by 0.3\%. 

\subsection{Model metal}
\begin{figure}[t]
\centering
\begin{minipage}[b]{.50\linewidth}
\includegraphics[width=\linewidth]{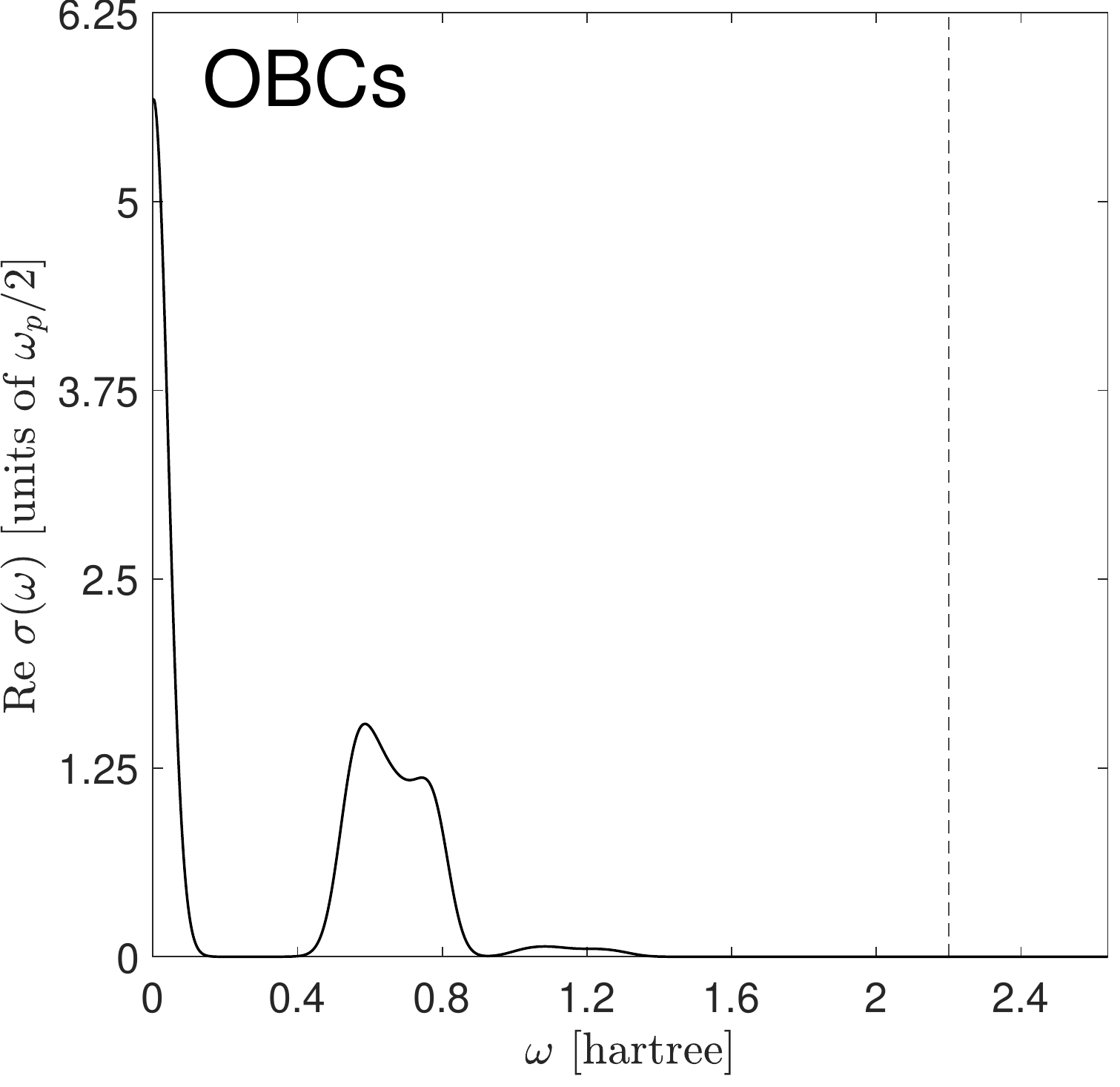}
\end{minipage}\hfill
\begin{minipage}[b]{.50\linewidth}
\includegraphics[width=\linewidth]{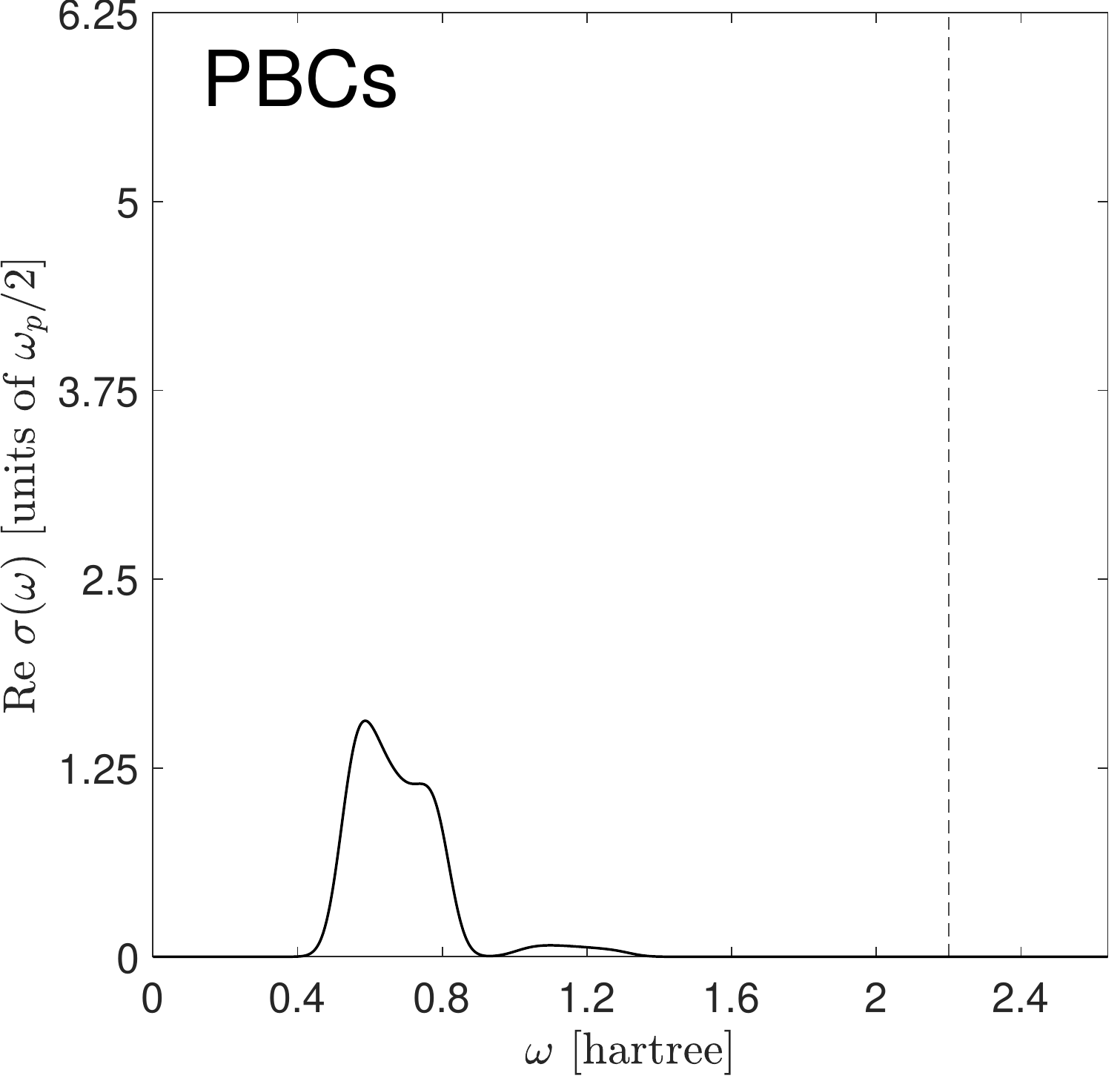}
\end{minipage}
\caption{Conductivity of the model metal in units of $\omega_{\rm p}/2$; the cutoff is shown as a vertical dashed line. Left panel: OBCs. Right panel: PBCs. Both plots are computed after \equ{Allen2}. The regular parts (and {\it only} the regular parts) almost coincide. Within PBCs \equ{Allen2} by itself does not account for the Drude~peak.
}
\label{fig:met} \end{figure}

We start showing in Fig.~\ref{fig:strength} the value of $D$ as a function of the periodic potential strength $V_0$, computed within PBCs, i.e.~with the the 1$d$ version of \equ{Allen1}. Starting from the free-electron $V_0=0$ case, $D$ decreases and converges to zero in the flat-band limit. All of the following simulations are performed at $V_0 = 0.8$, such that the spectral weights of the Drude and regular conductivities are comparable, with $D/2=0.57$.

Even in the metallic case we are close to completeness with a cutoff of 2.2 Ha: the $f$-sum rule is fulfilled at 99.97\% and 99.70\% within OBCs and PBCs, respectively. The conductivity plots evaluated from \equ{Allen2} in the two cases are shown in Fig.~\ref{fig:met}. As explained above, \equ{Allen2} within PBCs yields the regular (interband) term $\sigma^{(\rm regular)}(\omega)$ only; the Drude (intraband) term must be evaluated separately from \equ{Allen1}.

\begin{figure}[b] 
\includegraphics[width=\linewidth]{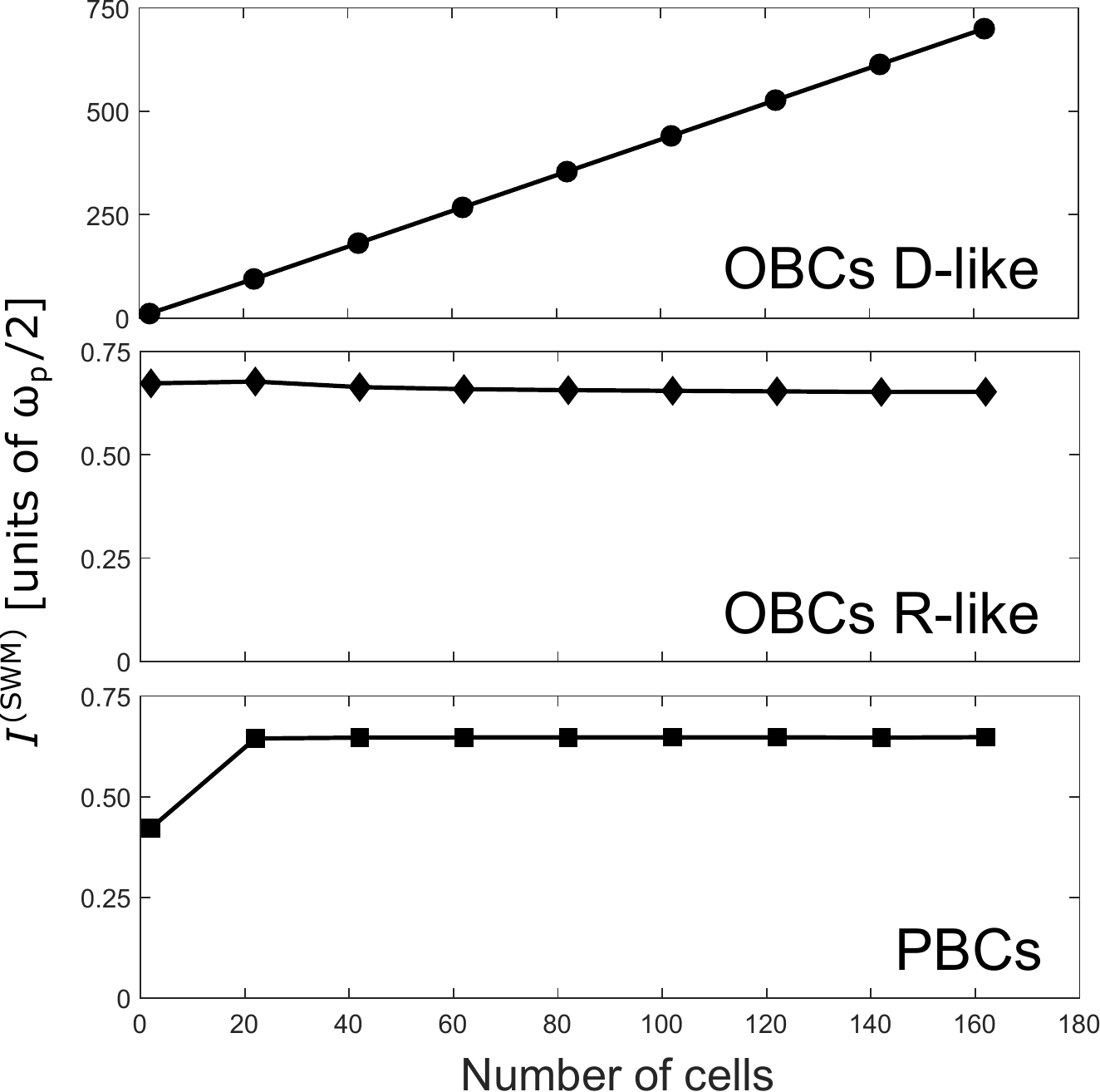} \\ 
\caption{Souza-Wilkens-Martin sum $I^{(\rm SWM)}$ as a function of the system size, computed from \equ{swm2}. In the OBCs case we show the separate contributions from the intraband (D-like) and interband (R-like) transitions; the former term diverges linearly with the system size, thus indicating a metallic state. In the PBCs case \equ{Allen2} accounts for the interband contributions only: $I^{(\rm SWM)}$ does not diverge.}
\label{fig:swm}
\end{figure} 

Within OBCs $\sigma(\omega)$, evaluated from \equ{Allen2}, saturates the $f$-sum. It shows two well-separated contributions, which clearly originate from the intraband (low-frequency) and interband (high-frequency) transitions. The spectral weight of the intraband transitions is accounted for, within PBCs, by the $\delta(\omega)$ Drude term. The same spectral weight is retrieved, within OBCs, in the low-frequency poles.
Previous considerations, based on the results in Fig.~\ref{fig:poles}, also show that such spectral weight accumulates at $\omega=0$ in the large-system limit. The low-frequency peak in the OBC conductivity is indeed the main focus of the present work; we are going to closely investigate it in the following. 

To start with, it is expedient to compare Fig.~\ref{fig:met} (left panel) to the free-electron case at the same density, Fig.~\ref{fig:sigma-free}. We clearly see that the effect of switching the periodic potential on is essentially a rescaling of the Drude peak: part of its spectral weight is transferred to the regular-conductivity term.

In the flat region between the two OBCs contributions the conductivity $\sigma(\omega)$ is (exponentially) vanishing. By choosing $\bar\omega$ in the middle of this region, we partition the OBCs $f$-sum  as: \[ \int_0^\infty d \omega \; \mbox{Re } \sigma(\omega) = \frac{\tilde D}{2} + \int_{\bar\omega}^\infty d \omega \; \mbox{Re } \sigma(\omega),  \] \[ \frac{\tilde D}{2} =  \frac{2 \pi e^2}{\hbar L^d} \sum_{\substack{\epsilon_n \leq \mu \\ \epsilon_m > \mu}} \; \sum_{\omega_{nm} < \bar\omega} \frac{|\me{n}{v}{m}|^2}{\omega_{nm}} . \label{tildeD} \] The value of $\tilde D$ obtained from \equ{tildeD} differs from the PBCs $D$ value, as from \equ{Allen1}, by 0.3\%. This major finding proves that---although a bounded sample does not support a dc current---the forced oscillations at low energy provide the quantitative value of $D$, which in turn is a measure of the (inverse) inertia of the many-electron system. 

Finally we address the SWM sum rule in the metallic case. Upon general grounds we expect that $I^{(\rm SWM)}$, evaluated from \equ{swm2} within OBCs, diverges linearly with the system size.\cite{rap_a33,rap156} Fig.~\ref{fig:swm} (top panel) confirms this behavior.
Within PBCs \equ{swm} is clearly nondivergent when evaluated using the interband term $\sigma^{(\rm regular)}(\omega)$ only in \equ{swm}: this is shown in the bottom panel of Fig.~\ref{fig:swm}. As explained elsewhere,\cite{Souza00,rap132,rap_a33} this is a geometrical property of the electronic ground state.

\section{Conclusions}

The real part of conductivity in bounded metallic systems within OBCs exhibits a qualitatively different behavior from that of analogous systems within PBCs.\cite{Akkermans97,Rigol08} Such difference stems from the response of the many-electron system to a dc field, which in metals induces free acceleration within PBCs but not within OBCs. Here we have thoroughly investigated the issue at the independent-electron level. The adiabatic inverse inertia of the electrons is measured by the Drude weight $D$, which has a well-known PBC expression (even beyond independent electrons\cite{Kohn64}), while instead it is formally zero within OBCs.

Upon general grounds, one expects that both kinds of boundary conditions should produce a given intensive observable in the thermodynamic limit. The apparent paradox has been previously solved in Ref. \onlinecite{Rigol08} in terms of lattice models at very high temperatures, which reduce finite-size effects; here instead we address band insulators and band metals in the conventional framework of zero-temperature electronic structure. Within PBCs the Drude weight at zero temperature originates from the adiabatic term in the Kubo formula (as shown in the Appendix); within OBCs it originates instead from the low-frequency sector of the nonzero-frequency Kubo formula. We have shown that the root of the difference is in the different selection rules for the intraband transitions.

Our 1$d$ simulations---with a model periodic potential---show how to actually evaluate $D$ to very high accuracy from the OBCs Kubo formula for conductivity; the role of the $f$-sum rule and of basis-set completeness is shown to be essential in order to control the numerical error.
Remarkably, our approach calls for a perspicuous physical interpretation: an oscillating low-frequency field induces---in a bounded metallic crystallite---forced oscillations, which are dominated by the many-electron inertia. The response carries therefore the same essential information as the response to a constant field within PBCs (i.e.~free acceleration).

The SWM sum rule provides, via a kind of fluctuation-dissipation theorem, a geometrical property of the electronic ground state which discriminates between insulators and 
metals.\cite{Souza00,rap132,rap_a33,rap156} This property has been ascribed so far to the mean-square quantum fluctuations of dc polarization, which are qualitatively different in insulators vs.~metals for bounded samples within OBCs. Here we have provided a complementary interpretation based on the OBCs Drude weight.

In conclusion, $D$ can be evaluated in two different ways: either by probing the free acceleration induced by a dc field (within PBCs), or probing the forced oscillations induced by a low-frequency field (within OBCs). We conjecture that such a general principle applies in general to any metallic many-electron system, well beyond the simple models thoroughly addressed in this work.

\vfill

\section*{Acknowledgments} 
 
This work has been supported by the ONR Grant No.~N00014-17-1-2803.

\vfill

\appendix*
\setcounter{equation}{0}
\section*{APPENDIX: RELATING KOHN'S APPROACH TO THE KUBO FORMULA}

According to Kohn\cite{Kohn64} the adiabatic limit requires performing the derivatives at a finite size, and the thermodynamic limit afterwards.\cite{Scalapino93} We therefore cast the current density in a band metal at double occupancy as  \[ \j  = - \frac{2 e}{L^d} \sum_{j\k} f_{j\k} \, {\bf v}_{j\k}  , \quad {\bf v}_{j\k} = \frac{1}{\hbar} \frac{\partial
{\epsilon}_{j\k}}{\partial \k} \label{a1} ,\] where the $\k$-point set is discrete, and the $T=0$ occupancies are $f_{j\k} = \theta(\mu-\epsilon_{j\k})$;
the dc conductivity is \[ \sigma_{\alpha\beta}^{(\rm D)} = \frac{\partial j_\alpha}{\partial E_\beta} =  - \frac{2 e}{\hbar L^d} \sum_{j\k} f_{j\k} \, \frac{\partial^2 \epsilon_{\k}}{\partial k_\alpha \partial E_\beta} . \] 
As explained in the main text, it is mandatory to adopt the vector-potential gauge, ergo 
 \[ \sigma_{\alpha\beta}^{(\rm D)}(\omega) =  - \frac{2 e}{\hbar L^d} \frac{d A(\omega)}{d E(\omega)} \sum_{j\k} f_{j\k} \,  \frac{\partial^2 \epsilon_{j\k}}{\partial k_\alpha \partial A_\beta} , \label{static} \] where  only the response of the many-electron system to a {\it static} vector potential ${\bf A}$ is considered;\cite{Scalapino93} therefore \equ{static} yields solely the adiabatic contribution to conductivity (the Drude term).

Given that $\E(\omega) = i\omega {\bf A}(\omega)/c$, causal inversion yields \[  \frac{dA(\omega)}{d E(\omega)} =  -c \left[ \pi \delta(\omega) + \frac{i}{\omega} \right] . \] The  perturbed band Hamiltonian is \[ H_\k = \emi{\k \cdot \r} H \ei{\k \cdot \r} = \frac{1}{2m} \left( {\bf p} + \hbar \k +\frac{e}{c} {\bf A} \right)^2 + V(\r) ; \] we thus exploit \[ \frac{\partial}{\partial {\bf A}} = \frac{e}{\hbar c} \frac{\partial}{\partial \k} , \] in order to retrieve \[ \sigma_{\alpha\beta}^{(\rm D)}(\omega) = \frac{2\pi e^2}{\hbar^2 L^d} \left[ \delta(\omega) + \frac{i}{\pi \omega} \right] \sum_{j\k} f_{j\k} \, \frac{\partial^2 \epsilon_{j\k}}{\partial k_\alpha \partial k_\beta} , \label{coinc} \] which is indeed---exploiting \equ{limit}---Kohn's Drude term in the case of a band metal. The physical interpretation of $\sigma_{\alpha\beta}^{(\rm D)}(\omega)$ is worth stressing: $D_{\alpha\beta}$ itself yields the current linearly induced by ${\bf A}$ (times a trivial factor),  while the $\omega$-dependent factor is the derivative of ${\bf A}$ with respect to~$\E$.\cite{Scalapino93}

In the present context the Kubo formula for $\sigma_{\alpha\beta}^{(\rm D)}(\omega)$ obtains straightforwardly from time-independent perturbation theory: it is enough to adopt the well-known effective-mass theorem\cite{em} in order to get the sum-over-states expression for $D_{\alpha\beta}$. Below we get some further insight by arriving at the same expression via an alternative path.

The periodic orbitals $\ket{u_{j\k}}= \emi{\k \cdot \r} \ket{\psi_{j\k}}$ are eigenstates of ${H}_\k$, hence the identity $\me{u_{j\q}}{( {H}_\k - \epsilon_{j\k})}{u_{j\q}} \equiv 0$ holds. Taking two derivatives, one arrives at \[  m^{-1}_{j,\alpha\beta}(\k) = \frac{1}{m} \delta_{\alpha\beta} - \frac{2}{\hbar^2} \mbox{Re } \me{\da u_{j\q}}{( {H}_\k - \epsilon_{j\k})}{\db u_{j\q}} , \] where in the presence of time-reversal symmetry the matrix element is actually real;
\bea  D_{\alpha\beta} &=& \pi e^2  \frac{n}{m}\delta_{\alpha\beta} - \frac{4 \pi e^2}{\hbar^2 L^d}  \sum_{j\k} f_{j\k} \,  \nn &\times&   \me{\da u_{j\q}}{( { H}_\k - \epsilon_{\k})}{\db u_{j\q}} ;  \label{geom} \eea therein the first term on the right-hand side is the free-electron Drude weight, while the second one is a ``geometrical'' correction,\cite{rap154}  accounting for the fact that the periodic potential hinders the acceleration of the free electrons. In a flat potential the $\ket{u_{j\k}}$ are $\k$-independent, ergo the correction vanishes.

We evaluate the $\k$-derivatives via perturbation theory:\\
\begin{widetext}
\[ \ket{\db u_{j\q}} = \sum_{j' \neq j} \ket{u_{j'\q}} \frac{\me{u_{j'\q}}{\db H_\k}{u_{j\q}}}{\epsilon_{j\k} -  \epsilon_{j'\k}} = \hbar \sum_{j' \neq j} \ket{u_{j'\q}} \frac{\me{u_{j'\q}}{v_{k_\beta}}{u_{j\q}}}{\epsilon_{j\k} -  \epsilon_{j'\k}} ; \] 
\[ \me{\da u_{j\q}}{({H}_\k - \epsilon_{j\k})}{\db u_{j\q}} = \hbar^2 \sum_{j' \neq j} \frac{\me{u_{j\q}}{v_{k_\alpha}}{u_{j'\q}} \me{u_{j'\q}}{v_{k_\beta}}{u_{j\q}}}{\epsilon_{j'\k} -  \epsilon_{j\k}} .\]
Replacing into \equ{geom}, and taking the thermodynamic limit as per \equ{limit},  one finally arrives at the sought for Kubo formula for $D_{\alpha\beta}$, \[ D_{\alpha\beta} = \pi e^2  \frac{n}{m}\delta_{\alpha\beta} - 4 \pi e^2 \sum_{j' \neq j} \intk f(\epsilon_{j\k})  \frac{\me{u_{j\q}}{v_{k_\alpha}}{u_{j'\q}} \me{u_{j'\q}}{v_{k_\beta}}{u_{j\q}}}{\epsilon_{j'\k} -  \epsilon_{j\k}} ; \label{kubo1} \]   a trivial transformation yields the more symmetric form,
\[ D_{\alpha\beta} = \pi e^2  \frac{n}{m}\delta_{\alpha\beta} - 2 \pi e^2 \sum_{j'\neq j} \intk \frac{f(\epsilon_{j\k}) - f(\epsilon_{j'\k})}{\epsilon_{j'\k} -  \epsilon_{j\k}}  \me{u_{j\q}}{v_{k_\alpha}}{u_{j'\q}} \me{u_{j'\q}}{v_{k_\beta}}{u_{j\q}}  \label{kubo2} . \] Finally it is worth noticing that the $f$-sum rule, \equ{fsum}, implies
\[   \int_0^\infty d \omega \; \mbox{Re } \sigma_{\alpha\beta}^{(\rm regular)} (\omega) = \pi e^2 \sum_{j'\neq j} \intk \frac{f(\epsilon_{j\k}) - f(\epsilon_{j'\k})}{\epsilon_{j'\k} -  \epsilon_{j\k}}  \me{u_{j\q}}{v_{k_\alpha}}{u_{j'\q}} \me{u_{j'\q}}{v_{k_\beta}}{u_{j\q}} .\]
\end{widetext}

\begin{thebibliography}{10}

\bibitem{Vanderbilt}
{ D.~Vanderbilt, {\it Berry Phases in Electronic Structure Theory} (Cambridge
  University Press, Cambridge, 2018)}.

\bibitem{Kohn64}
{ W.~Kohn, Phys. Rev. {\bf 133}, {A171} (1964)}.

\bibitem{Kohn68}
{ W.~Kohn, in {\it Many--Body Physics}, edited by C. DeWitt and R. Balian
  (Gordon and Breach, New York, 1968), p. 351}.

\bibitem{Shastry90}
{ B.~S.~Shastry and B.~Sutherland, Phys.~Rev.~Lett. {\bf 65}, 243 (1990)}.

\bibitem{Scalapino93}
{ D.~J.~Scalapino, S.~R.~White, and S.~C.~Zhang, Phys. Rev.~B {\bf 47}, 7995
  (1993)}.

\bibitem{Allen06}
{ P.~B.~Allen, in {\it Conceptual foundations of materials: A standard model
  for ground- and excited-state properties}, edited by S.~G.~Louie and M.~L.~Cohen (Elsevier, Amsterdam, 2006), p.~139}.

\bibitem{Rigol08}
{ M.~Rigol and B.~S.~Shastry, Phys.~Rev.~B {\bf 77}, 161101(R) (2008)}.

\bigskip

\bibitem{Souza00}
{ I.~Souza, T.~Wilkens, and R.~M.~Martin, Phys.~Rev.~B {\bf 62}, 1666 (2000)}.

\bibitem{rap157}
{ R.~Resta, J.~Phys.~Condens. Matter {\bf 30}, 414001 (2018)}.

\bibitem{Drude00}
{ { P.~Drude, Annalen der Physik. {\bf 306}, 566 (1900)}}.

\bibitem{AM1}
{ { N.~W.~Ashcroft and N.~D.`Mermin, {\it Solid State Physics} (Saunders,
  Philadelphia, 1976)}}.

\bibitem{AM-Ch13}
{See Chap.~13 in Ref.~\onlinecite{AM1}}.

\bibitem{nota}
{ The ${\bf q} \rightarrow 0$ (infinite wavelength) and $\omega \rightarrow 0$
  (dc) limits in general do not commute. The consequences of this in the case
  of $D$ are thoroughly discussed in Ref. \onlinecite{Scalapino93}}.

\bibitem{rap100}
{ R.~Resta, Phys.~Rev.~Lett. {\bf 80}, 1800 (1998)}.

\bibitem{Akkermans97}
{ E.~Akkermans, J.~Math.~Phys. {\bf 38}, 1781 (1997)}.

\bibitem{rap132}
{ R.~Resta, J.~Chem.~Phys. {\bf 124}, 104104 (2006)}.

\bibitem{rap_a33}
{ R.~Resta, Riv.~Nuovo Cimento {\bf 41}, 463 (2018)}.

\bibitem{rap156}
{ A.~Marrazzo and R.~Resta, Phys.~Rev.~Lett. {\bf 122}, 166602 (2019)}.

\bibitem{em}
{ See Appendix E in Ref.~\onlinecite{AM1}}.

\bibitem{rap154}
{ R.~Resta, arXiv:1703.00712, rejected by Phys.~Rev.~Lett.}

\end{thebibliography}

\end{document}